\documentstyle[aps,multicol,epsfig]{revtex}   
\def\bea{\begin{eqnarray}}
\def\eea{\end{eqnarray}}
\def\be{\begin{equation}}
\def\ee{\end{equation}}
\begin{document}
%\tightenlines
\draft
\author{Bogdan Damski$^1$, Zbyszek P. Karkuszewski$^{1,2}$, Krzysztof Sacha$^1$, 
and Jakub Zakrzewski$^1$,}
\address{
$^1$Instytut Fizyki imienia Mariana Smoluchowskiego,
  Uniwersytet Jagiello\'nski,\\
 ulica Reymonta 4, PL-30-059 Krak\'ow, Poland \\
        $^2$Theoretical Division, T6, MS B288,\\
        Los Alamos National Laboratory,
        Los Alamos, NM 87545 USA\\
}
\title{Simple method for excitation of a Bose-Einstein condensate}
\date{\today}
\maketitle
\begin{abstract}
An appropriate, time-dependent
modification of the trapping potential may be sufficient to
create effectively collective excitations
in a cold atom Bose-Einstein condensate. The proposed
method is complementary to earlier suggestions and should allow 
the creation of both dark solitons and vortices.
\end{abstract}
\begin{multicols}{2}

\section{Introduction}

Soon after the first
spectacular realizations of the Bose-Einstein condensate
(BEC) in cooled and trapped atomic gases
 \cite{cornell95,ketterle95,bradley97}, investigations
 of possible new effects involving the condensate appeared.
The BEC allows us to study several typical quantum mechanical phenomena
 on a macroscopic level -- because the macroscopic sample of atoms is
 described by a single wavefunction.
A standard  example is the
splitting of the condensate into two spatially separated parts \cite{ketterle97} 
followed by a superposition of the parts.
The observation of the
interference fringes \cite{ketterle97,zoller96,wallis97} is a manifestation of
the quantum coherence between two macroscopic parts of the condensate.
By leaking the atoms from the condensate (typically downwards -- due to gravity)
one may prepare an ``atom laser'' \cite{zoller96a,ketterle97a}.

It has been also realized that
collisions between spatially separated condensates may be used to
create collective excitations in the condensate
 \cite{reinhardt97,scott98}. Assume a standard 
  mean field single
particle description of the gas of weakly interacting bosons in the
limit of vanishing temperature (for reviews see
 \cite{parkins98,dalfovo99,leggett01}). The time-dependent
 equation governing the state of the BEC is
 then the celebrated  Gross-Pitaevskii equation (GPE).
 Its solutions may describe either solitary waves or vortices as is
typical for the nonlinear equation.
For characterization of solitons the language of nonlinear optics
\cite{optics} is quite useful. One may consider then
bright solitons (bell shaped
structures propagating without dispersion), dark solitons (with
a node in the middle -- an analog of the first excited state in the
non-interacting particles picture) or the intermediate grey
solitons.

Not only collisions may be used to create excitations of the BEC.
In fact several schemes have been proposed, some of them being successfully
applied in experiments. It has been suggested that
a resonant Raman excitation scheme may be utilized
to excite vortex states\cite{marzlin97}. 
However, the resonance is
modified appreciably
during the process of transferring the population from the ground
state to the vortex state due to the nonlinearity of the GPE.
An apparently more robust approach is the adiabatic scheme of \cite{dum98}
 which takes full account of the nonlinearity. 
 It
utilizes a controlled laser induced adiabatic transfer, populating
 solitonic or vortex solutions
of GPE, depending on the details of the excitation. The adiabatic
transfer uses internal atomic
transitions combined with appropriate states of the condensate.
A phase imprinting method, originally proposed in \cite{dobrek99},
produces a phase shift between two parts of the condensate. This
method in fact
has been applied experimentally to create dark
  solitons both in cigar
shaped BEC \cite{burger99} and in the spherically symmetric condensate
  \cite{denschlag00}.

Another technique
based on laser stirring of the condensate allows the production of several
different vortex states \cite{madison00}. Recently the method has been applied
to create lattices containing over 100 vortices \cite{ketterle01}.

The aim of this paper is to discuss in detail  yet another method which,
in our opinion, may serve to generate
collective excitations in a BEC. It allows for  the  creation of
grey (or even dark) solitons as well as vortices.
 The method which we propose resembles to a certain extent the
 adiabatic passage scheme of \cite{dum98}. In the latter, the
transfer of population between two internal atomic states is
accompanied by an appropriate change of the condensate wavefunction
into a dark soliton, two-soliton or vortex solution of the GPE
\cite{dum98}.
Our method originally proposed for non-interacting
particles \cite{kark01} and extended to weakly attractive interaction 
 assumes a fast sweep of
the laser beam across the trap. In this way
the trapping potential varies with time enabling a transfer
of population to excited BEC states. Under appropriate conditions
this approach allows for an efficient creation of collective
excitations in the condensate as explained below.

In Section II we present the
method for a non-interacting particles model. We discuss both the
excitation of solitons
 in the one dimensional (1D) case
\cite{kark01} as well as the possibility of vortex creation in
the effective two-dimensional (2D) example.
 In the next sections we extend the approach to interacting
particles. In Section III
  we apply it to
particles with  repulsive atom-atom interactions
 (positive scattering length $a_0$)
in 1D model case. In this way we complement our earlier study
for attractive interactions \cite{kark01} for the case most
often met in experiments with a BEC \cite{cornell95,ketterle95}.
Extending the treatment to 2D, we  also consider  the
 excitation of vortices for interacting particles.
In Section IV we show some related observations on behavior of GPE
energy levels  while changing a parameter of the potential.

\section{Non-interacting particles}
Let us first consider the simplest situation: The case of
 non-interacting particles. A BEC of non-interacting particles
is not realized in nature, yet it may serve as
a good model to describe the basic idea underlying
the scheme we proposed. For  an excitation
of solitons propagating along a given direction 1D model is clearly
sufficient. For excitation of vortices, though,  at least a
2D model is required. Such low-dimensionality models may
be fully justified for non-interacting particles, assuming
separability of the trapping potential. For
interacting particles such simplified models may be of value
for appropriately prepared (cigar shaped or flat disc shaped,
respectively) condensates 
\cite{parkins98,dalfovo99,leggett01,garcia98,jackson98,muryshev99,fedichev99,busch00}.

Let us discuss the excitation of solitons first.
Consider the condensate which occupies the ground state of the
harmonic trap.
Let us now sweep the region where the condensate is located with
an additional laser beam, whose frequency is
 appropriately tuned close to
the resonance of some internal
atomic transition. By an adiabatic elimination of the upper
atomic state (possible if the laser is detuned
from the exact resonance) one may show \cite{graham92} that
 such a laser beam creates an effective
additional potential well (or a barrier, depending on the sign of the
detuning with respect to the atomic transition) for the
motion in the external atomic degree of freedom. The laser intensity is
typically Gaussian-shaped in the direction perpendicular to
the direction of propagation. The atoms experience a potential proportional
to the intensity which may be
represented as 
\begin{equation}
V(x)=\frac{x^2}{2}+U_0\arctan(x_0)
\exp\left(\frac{-(x-x_0)^2}{2\sigma^2}\right).
\label{poten}
\end{equation}
We use the trapping harmonic oscillator units, i.e.
$\omega t$ for time and $\sqrt{\hbar/m\omega}$ for length, where $\omega$ is
harmonic oscillator frequency while $m$ stands for atomic mass.
(These units are also used through the rest of the paper). A similar 
modification of the potential has been used already in experiments to
split the condensate into two parts \cite{ketterle97}.
We propose modifying the trapping
potential in a different way. Assume that the local Gaussian well is
created on the very edge of the harmonic potential well thus not affecting 
the condensate. Then we change the laser
beam direction slowly, in effect sweeping the well across the
trapping potential. At the same time we gradually decrease
the laser intensity, thereby decreasing
the depth of the well. Such a procedure is equivalent to a change
 of $x_0$ from some negative
value to zero in (\ref{poten}) assuming $U_0$ positive.

\begin{figure}
\centering
{\epsfig{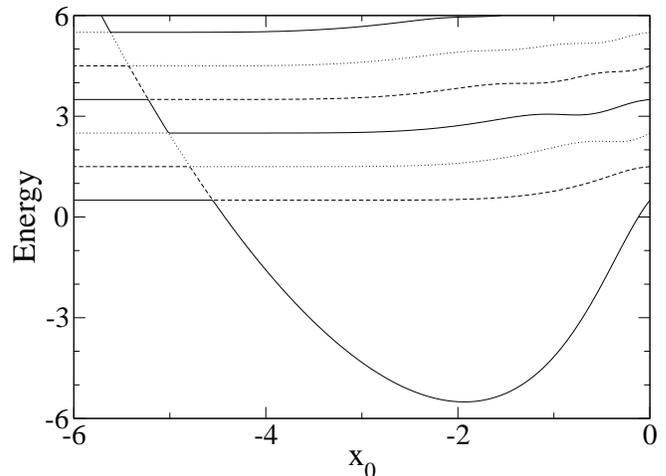}}
\caption{
Energy levels for a single particle in the potential
({\protect \ref{poten}}) for $U_0=13.4$ and $\sigma=0.2$ as a function of $x_0$.
Note the narrow avoided crossing between the ground (the lowest solid line) 
and first excited (the lowest dashed line) states around $x_0=-4.5$. Similar
avoided crossings occur between the first and second excited states, between
the second and third ones and so on.
}
\label{levdyn1}
\end{figure}

The procedure proposed has a clear quantum mechanical meaning.
 For a sufficiently slow change of the potential  the system adiabatically
 follows its quantum energy levels
{\it except} in the vicinity of avoided crossings. The energy gap of crossings
may be controlled by choosing appropriate values of $U_0$
and $\sigma$ in (\ref{poten}). In particular it is easy to arrange
that the avoided crossing between the ground and the first excited
state of the potential occurs during the sweep of the local potential
well (see Fig.~\ref{levdyn1}). Moreover such an avoided crossing may be made sufficiently narrow
to be passed {\it diabatically} during the potential sweep. Then,
when the local potential well disappears the particle, originally
in the ground state of the harmonic trap, is left with a high probability, $p_1$,
in a first excited state (the Landau-Zener transition).
The efficiency of the process depends on the size of the avoided crossing
(which should be much smaller than the mean splitting between levels)
and how quickly the potential is modified.

As discussed in the earlier report \cite{kark01}, numerical
simulations fully confirm the proposed scheme.
Without any special optimization attempt, choosing
the parameters of the potential (\ref{poten}) as $U_0=13.4$, $\sigma=0.2$ and
changing $x_0$ from $-7$ to 0 with the velocity $\dot x_0=0.02$
yields $p_1=0.99$. Similar values of $p_1$ are obtained for different values
of $U_0$ and $\sigma$. The method is also robust with respect to
the functional form of the potential well. We have checked
that similar $p_1$ values are obtained if instead of the $\arctan (x_0)$
we use other smooth monotonically changing functions of $x_0$.

Looking at the level dynamics as a function of $x_0$
(compare Fig.~\ref{levdyn1})  one readily
realizes that it is easy to
generalize the excitation mechanism to obtain higher
excitations of the condensate. It is sufficient to sweep the local
potential well twice in order to get a highly efficient transfer of
population from the ground state to the second excited state of the
potential.  This extension is easily tested numerically in the model with
non-interacting particles. Keeping the same parameters as in the
single excitation case but merely repeating the potential sweep the second
time we get $p_2=0.99$ as the squared overlap between the final wavefunction
and the second excited state of the trap. 
A third consecutive sweep yields a ``triply'' exited state with the
probability $p_3=0.99$ again without any modification of the parameters
of the sweeping potential. 

All these tests of single as well as multiple excitations
indicate that, while we propose to excite
the condensate by sweeping the trapping
potential using the local potential well, the excitation
may in fact be realized in a number of different ways. The key feature 
necessary for our method is the presence a narrow isolated avoided crossing between
the ground and excited states.

\begin{figure}
\centering
{\epsfig{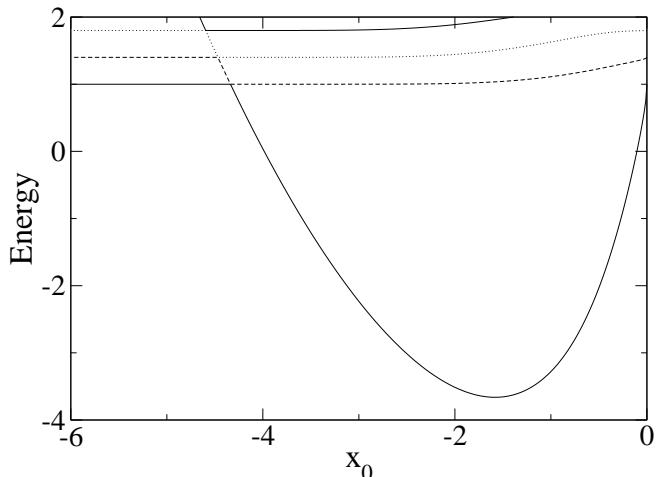}}
\caption{
Energy levels, in the rotating frame ($\Omega=0.6$), for a single particle in 
the potential ({\protect \ref{pot2D}}), for $U_0=25$ and $\sigma=0.2$, as 
a function of $x_0$.
Note the narrow avoided crossing between the ground (the lowest solid line) 
and first excited (the dashed line) states around $x_0=-4.5$. 
For $x_0=0$ the energy levels correspond to 2D harmonic oscillator states 
with $L_z=0$, $L_z=1$, $L_z=2$ and $L_z=3$ from bottom to top. 
}
\label{levdyn2d}
\end{figure}
Our method to create vortices in a 2D model system
is a natural extension of the former approach. It consists of
diabatic transition through an isolated avoided crossing.
Now we consider
a BEC in a cylindrically symmetric 2D harmonic trap. An additional
laser beam producing a local potential well is now 
rotating with frequency $\Omega$ around the symmetry axis of the trapping
potential.
However, the energy is not conserved: Looking at
the system in the frame rotating with a frequency
$\Omega$ 
we may consider the eigenvalues of the Hamiltonian operator
\begin{equation}
H=-\frac{1}{2}\left(\frac{\partial^2}{\partial x^2}+
\frac{\partial^2}{\partial y^2}\right)
+U(x,y)-\Omega L_z,
\label{hrot}
\end{equation}
where
\begin{eqnarray}
U(x,y)&=&\frac{x^2+y^2}{2}- \cr
&&U_0\sqrt{\arctan(|x_0|)} 
\exp\left(\frac{-(x-x_0)^2+y^2}{2\sigma^2}\right).
\label{pot2D}
\end{eqnarray}
For different distances $|x_0|$ of the laser beam from the center of the
trapping potential we can calculate energy levels of the Hamiltonian (\ref{hrot}).
For appropriate values of $U_0$, $\sigma$ and $\Omega$, when changing 
$x_0$ from some negative value to zero, one can observe
a narrow avoided crossing between the ground and first excited states 
(see Fig.~\ref{levdyn2d}). 
The latter corresponds, for vanishing laser beam, to the first excited 
state of the harmonic potential with $L_z=1$. 

For the same
parameters as used in Fig.~\ref{levdyn2d}, we perform a time dependent 
numerical simulation.
Starting from a ground state of the condensate and changing $x_0$ from $-5$ to 0
with a velocity $\dot x_0=0.036$, we obtain the
first excited state with $L_z=1$ with more than 99\% accuracy.
One may
envision that after a single sweep we illuminate a condensate
second time with a similar sweep. This, via a second
avoided crossing (compare Fig.~\ref{levdyn2d}) yields $L_z=2$
state with high efficiency (more than 99\%).

Provided, therefore, that the resulting picture is not modified
strongly by the interaction of particles constituting a BEC
the proposed method should be able to produce both solitons
and vortices with quite high efficiency. The next sections
describe the effect of the interactions on the proposed mechanism.

\section{Collective excitations in BEC of interacting particles}

While the proposed method seems to be 
quite robust for non-interacting particles
model its applicability to a BEC of interacting particles is far from
clear. After all, the particle interaction  necessarily changes
the energy levels of the system. The proposed method relies on
narrow avoided crossings between levels when changing the parameter
of the system -- it is thus sensitive to the details of level dynamics.
It is not obvious whether the presence of the interaction between the
particles will not destroy the proposed mechanism of the excitation.

The partial preliminary answer has been given 
already in \cite{kark01}, where we considered the effect of attractive
atom-atom interactions present in a BEC of
 Li atoms \cite{bradley97} on the creation of solitons in a 1D model.
For such a condensate the number of particles is
not too big and the effect of atom-atom interactions on the
behavior of the system is rather small. By a direct integration of 
the time-dependent GPE in 1D
\begin{equation}
i\frac{\partial \psi}{\partial t}=-\frac{1}{2}\frac{\partial^2 \psi}{\partial x^2}
+V(x)\psi+g|\psi|^2\psi,
\label{gpe}
\end{equation}
(starting with the ground state
of the condensate in the harmonic trap for $g=-5$) we
were able to get a 97.5\% population transfer into a
collective state corresponding to the first excited state
in the independent particle model. In (\ref{gpe}) $g$ is a measure of
nonlinearity and is proportional to number of particles $N$ in the BEC.
The  value $g=-5$ taken for numerical simulation corresponds, 
for a trap used in \cite{bradley97}, to $N=900$ atoms in the condensate 
fraction -- a typical number in Li condensate \cite{bradley97}.

\begin{figure}
\centering
{\epsfig{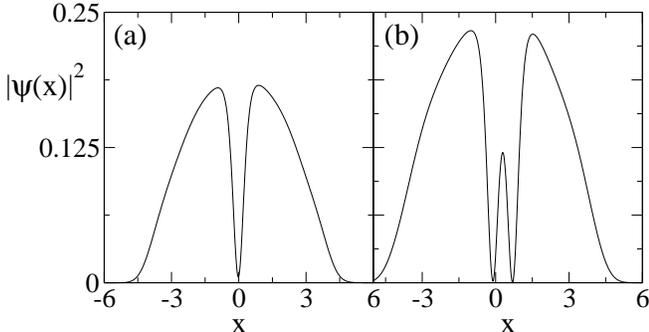}}
\caption{
Single particle reduced probability densities of the condensate
(i.e. solutions of the time-dependent GPE ({\protect \ref{gpe}}) for $g=50$) 
after the first 
[panel (a)] and second [panel (b)] potential sweeping.} 
\label{wprep1}
\end{figure}
Clearly this is not the full story. The 1D approach
for interacting atoms is not exact, since nonlinearity couples different
degrees of freedom. Still, a one-dimensional approach based on the GPE is often
used and may be justified for asymmetric traps
\cite{parkins98,dalfovo99,leggett01,garcia98,jackson98,muryshev99,fedichev99,busch00}. 
More importantly most of the condensates realized in laboratories
consist of particles with repulsive atom-atom interactions. Such condensates
can easily hold about $N=10^5$ particles 
\cite{parkins98,dalfovo99,leggett01}. Consequently the
nonlinear term in (\ref{gpe}) becomes much more
important than for the attractive interaction.
Thus the true test of our method requires a simulation for large positive
$g$ values.

Fig.~\ref{wprep1}a shows the final wavefunction obtained for the $g=50$ case
by a numerical
integration of Eq.~(\ref{gpe}), taking as the initial state the
ground state of the condensate for the same value of $g$. The parameters of the
potential are $U_0=13.4$ and $\sigma=0.2$, and the velocity of the sweep is
$\dot x_0=0.6$. The overlap
of the wavefunction depicted in Fig.~\ref{wprep1}a with the ideal ``excited''
state of the condensate  
is $p_1=0.98$ at the end of the potential sweeping. 

Similarly successful is a double application of the potential sweep
in order to obtain a ``two-node'' collective state of the condensate.
We use the same parameters as above. In fact for a second sweep we
start from the wavefunction shown in Fig.~\ref{wprep1}a and make
the second sweep identical to the first one. The results are depicted in
Fig.~\ref{wprep1}b. The  
overlap of the wavefunction at the end of the sweeping
with the exact solution of the time-independent GPE is $p_2=0.82$.

Consider now the excitation of vortices in the 2D model. 
The time dependent GPE in  rotating $x,y$ coordinates reads
\begin{equation}
i\frac{\partial \psi}{\partial t}=
H\psi
+g|\psi|^2\psi,
\label{hrot1}
\end{equation}
where $H$ is given by Eq.~(\ref{hrot}).
We get the time independent version  by substituting $\psi(x,y,t)=\exp(-i\mu t)
\varphi(x,y)$ with $\mu$ being the chemical
potential. The resulting time-independent equation 
\begin{equation}
H\varphi+g|\varphi|^2\varphi=\mu\varphi
\label{gpe3}
\end{equation}
\begin{figure}
\centering
{\epsfig{file=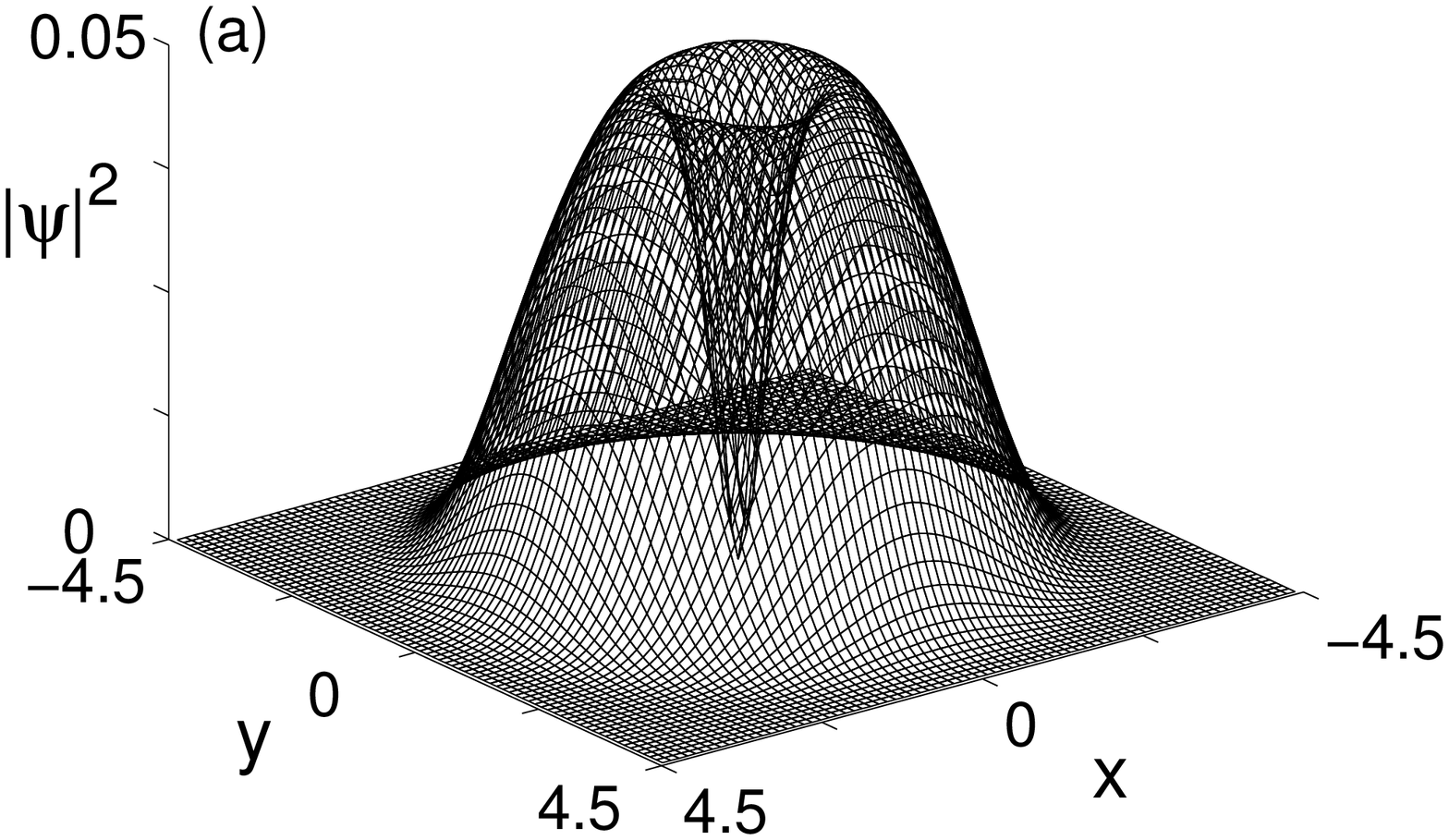, clip = true, width=8.6cm}}
{\epsfig{file=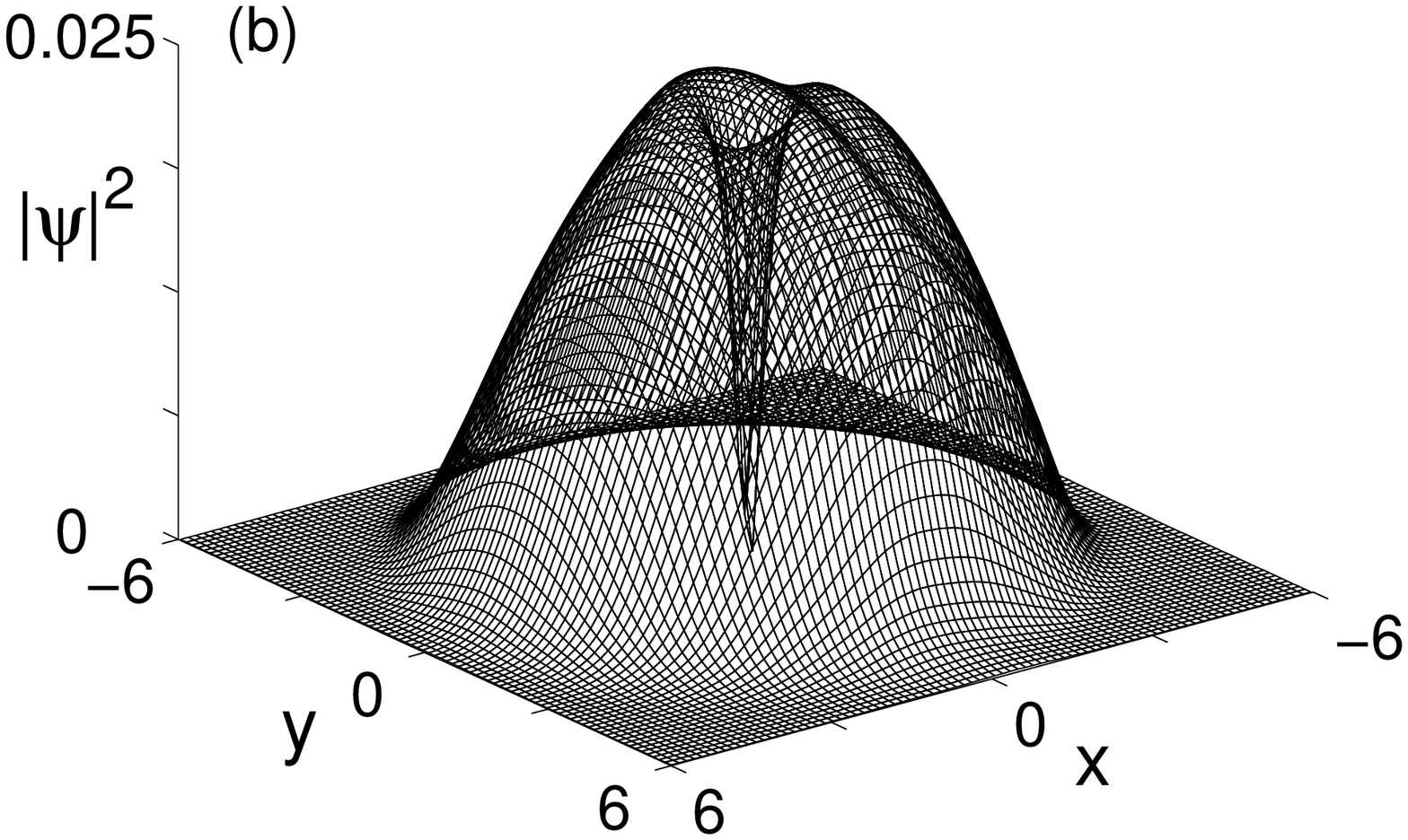, clip = true, width=8.6cm}}
\caption{
Panel (a):
single particle reduced probability density of the condensate
(i.e. solution of the time-dependent GPE ({\protect \ref{hrot1}}) for 
$g=100$, $U_0=25$, $\sigma=0.2$ and $\Omega=0.23$)
at the end of the potential sweeping 
 where $x_0$ has been changed from $-7$ to 0 
with a velocity $\dot x_0=0.35$, see ({\protect \ref{pot2D}}). 
Panel~(b): the same as in the panel (a) but for 
$g=500$, $\Omega=0.12$ and  $x_0$ going from $-9$ to 0 with a velocity of 0.53.
 }
\label{cont}
\end{figure}
\noindent
can be solved by the method of self-consistent field or by using the
imaginary time propagation approach. The former is quite robust for the 1D 
GPE but for 2D case it does not work efficiently. The latter is extremely
effective for the ``ground state'' of the condensate, however, its
application to ``excited'' states becomes difficult 
(especially for a trap without rotational symmetry) since for $g\ne 0$
the stationary solutions are no longer orthogonal. The third possibility
is to find the solution of Eq.~(\ref{gpe3}) by minimizing 
$\langle\phi|\phi\rangle$, where
\begin{equation}
\phi=H\tilde\varphi+g|\tilde\varphi|^2\tilde\varphi-\tilde\mu\tilde\varphi,
\end{equation}
with 
\begin{equation}
\tilde\mu=\langle\tilde\varphi|H+g|\tilde\varphi|^2|\tilde\varphi\rangle.
\end{equation}
Indeed, starting with some initial function $\tilde\varphi$ decomposed 
in a given basis, standard procedures of minimizing of multidimensional 
functions can lead to a desired solution provided the initial guess function 
$\tilde\varphi$ is sufficiently close to the exact solution $\varphi$.

With solutions of the time-independent GPE at hand, we may prepare
a BEC in its ground state, integrate the time-dependent GPE with
the spiral-like potential sweep (given by (\ref{pot2D}) in the rotating
frame),
and compare the final wavefunction with the excited vortex-like solutions
of the time-independent GPE. The resulting wavefunctions are presented
in Fig.~\ref{cont} for $g=100$ and $g=500$.
Their overlaps with the excited state corresponding to one quantum of the
angular momentum (per particle) are 0.99 and 0.98 respectively. The non-ideal 
population transfer is responsible for a slightly asymmetric shape
of the final wavefunctions. However, the method may
be quite effective in generating vortex-like excitations in a BEC.
Let us note that our approach resembles to some extent the stirring
approach to vortex creation \cite{madison00}. In that method the potential
as a whole is rotated with a certain frequency $\Omega$. In our approach
a static, cylindrically symmetric potential is supplemented, by an
additional laser beam, with a narrow (with respect to the BEC dimension)
structure moving along the spiral. Thus the physical picture of transferring
the angular momentum to the condensate is a bit different in both cases.
Naturally it differs also from the phase imprinting technique
\cite{dobrek99}.

\section{Behavior of levels for interacting particles}

\begin{figure}
\centering
{\epsfig{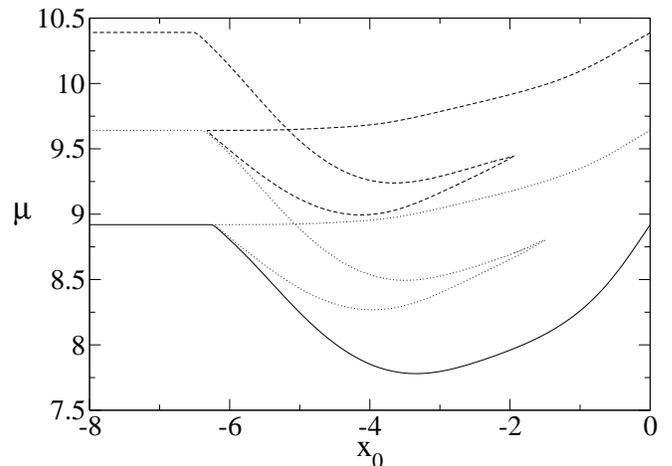}}
\caption{
Energy (i.e. chemical potential {\protect \cite{foot1}}) 
levels of the condensate
in the potential ({\protect \ref{poten}}) for $g=50$, $U_0=13.4$ and $\sigma=0.2$ 
versus a value of the parameter $x_0$, i.e. the position of the laser beam.
}
\label{loopsg}
\end{figure}

The original physical picture of the excitation scheme, 
as discussed in Section II for the 
non-interacting particles model, is based on the diabatic transitions 
between levels via narrow avoided crossings. The actual implementation 
for interacting particles has been tested by numerical integration of the
time-dependent GPE without resorting to the explicit changes of GPE 
levels with respect to a modification of the potential.
To see whether the same picture may be invoked for the interacting
particles we have solved the 1D time-independent GPE not only for the
harmonic potential but also in the presence of the laser beam.
Obtained energy levels \cite{foot1} are shown in Fig.~\ref{loopsg} as a 
function of the position of the center of the laser beam. Observe
the presence of loop-like structures. 
Such a behavior is impossible in the case of a linear Schr\"odinger equation.
Actually, it can be shown that the changes of levels' energies
 may be considered as a true Hamiltonian classical
dynamics where the energies of levels play the role of positions of
fictitious particles while the changing parameter corresponds to a
fictitious time \cite{pechukas83,yukawa85,haake,stockmann}.

\begin{figure}
\centering
{\epsfig{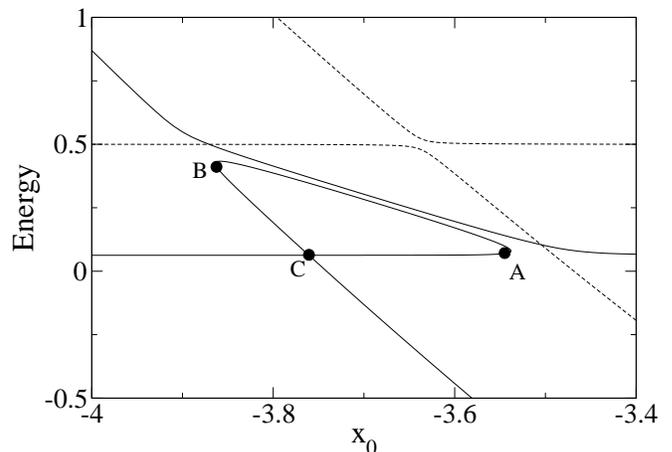}}
\caption{
Solid lines: energy 
(i.e. chemical potential {\protect \cite{foot1}}) levels of the
condensate in the potential ({\protect \ref{poten}}) for $g=-1$, $U_0=6.4$ 
and $\sigma=0.5$ versus a value of the parameter $x_0$, i.e. the position of 
the laser beam. Specific points of the ground state level are indicated by
letters, see text. Dashed lines are energy levels of the corresponding  
linear Schr\"odinger equation.
}
\label{smallg}
\end{figure}
To explain the appearance of the loops let us 
consider solutions of
the time-independent GPE for weak attractive particle
interactions. Figure \ref{smallg} shows the lowest energy levels 
for both the non-interacting and interacting case with $g=-1$
 [see (\ref{gpe3})].

In the non-interacting case  avoided crossings result from
changing the shape of the double well potential. To 
approximately predict energy levels in a double well one may consider each well
separately. Then for an asymmetric potential, 
the resulting energies in each well are usually considerably
different. However, for certain shapes, the energies become degenerate. 
Taking the tunneling between the wells into account, the level
crossing changes into an avoided crossing (see Fig.~\ref{smallg}). 
In the $g=0$ case, the change of the ground state energy  
(when going from the left to the right in Fig.~\ref{smallg}) corresponds 
to the transfer of probability density from the harmonic well to the local well 
created by a laser. At the point of the avoided crossing both wells are equally
populated i.e. the ground state consists of equally weighted symmetric 
superposition of eigenstates of the right and left well.

\begin{figure}
\centering
{\epsfig{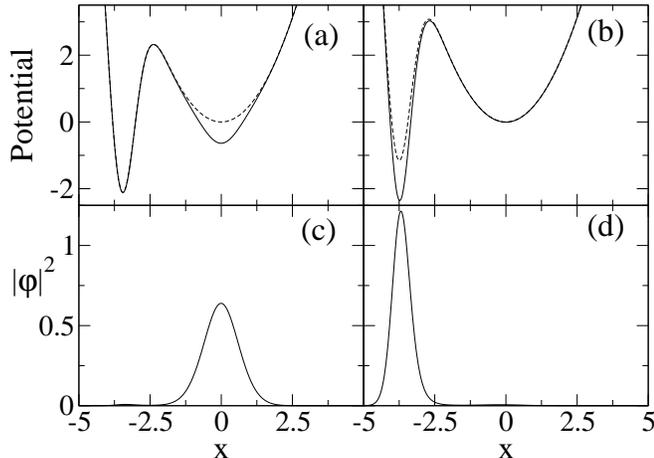}}
\caption{
Panel~(a): plots of the potential ({\protect \ref{poten}}) (dashed line)
and the effective potential  ({\protect \ref{ve}}) (solid line)
corresponding to point
A in Fig.~{\protect \ref{smallg}}. Panel (c): probability density of the
condensate at the point A in Fig.~{\protect \ref{smallg}}. Panel (b) and (d)
are the same as in the corresponding previous panels but for point B in 
Fig.~{\protect \ref{smallg}}.
}
\label{veff}
\end{figure}
Analyzing the interacting particle case we may consider a solution of 
the GPE as if it is a solution of the Schr\"odinger equation with an effective 
potential 
\begin{equation}
V_{eff}(x)=V(x)+g|\varphi(x)|^2.
\label{ve}
\end{equation}
For $g<0$, the interaction term deepens a potential well in which 
the probability density is localized (see Fig.~\ref{veff}). 
Consider the ground state of the harmonic trap and 
suppose that the laser beam approaches the center of this trap. 
At each position of 
the beam $x_0$, we can calculate approximate energies considering each well of 
the effective potential (\ref{ve}) separately. 
The resulting energy in the left well is
higher than the one in the right well when 
the probability density is situated in the right well (compare Fig.~\ref{veff}).
 This holds up to the point indicated as A in Fig.~\ref{smallg}, 
 where the lowest eigenenergy of left well treated alone approaches the energy 
 of the condensate in the harmonic trap. At that point we can
start populating the left well. However, transferring particles from the right 
to the left well breaks the energy balance between the two because of 
the density term
in the effective potential (\ref{ve}).  
The only way to restore balance is to lift the left well a little, i.e. to  
shift the center of the laser beam back to the left.

Transfer of particles from the harmonic trap to the dip accompanied by
shifting of the laser to the left can be carried on until nothing is left
in the former. When this stage is reached (point B in Fig.~\ref{smallg}),
all particles are localized in the left dip. Now this dip can be lowered
without populating the right well. This will
decrease the energy of the condensate
 which at some point matches the value
of the starting energy with all particles in the right well.
 This is
indicated by C in Fig.\ref{smallg}. 
The situation where it is possible to have the same values of a chemical 
potential for states with all particles in the left  or right well is in fact
a necessary condition for loop-like structure to be present.
Recall that in the non-interacting case, matching of corresponding energy 
levels is responsible for an avoided crossing.

For $g>0$ it is possible to have a similar scenario to the attractive 
particles case. 
The only exception is that the ground energy level does not reveal any loop (see
Fig.~\ref{loopsg}). Indeed, to transfer the probability density of the ground
state from the harmonic well to the well created by the laser beam we
have to move the beam further and further towards the center because, contrary
to the attractive particles case, taking the
density from the right well leads to a deepening of the effective potential 
(\ref{ve}) at that place.

Considering the strange shapes of levels in Fig.~\ref{loopsg} one may
wonder why we have been so successful with time-dependent numerical
approach. 
At the beginning of the excitation process, the parameter of the system is changed 
sufficiently slowly to follow
``adiabatically'' a single level far from a loop (``avoided crossing'')
region. In that region, on the other hand, we are diabatic in a sense that
there is insufficient time for the wavefunction to change its shape
appreciably. Thus for non-interacting particles we make a diabatic
jump from one branch of the avoided crossing to the other (following the
wavefunction).  Similarly, for interacting particles, loops are passed 
quickly in a way which locally minimizes
any significant change of the wavefunction shape.

To express this more quantitatively we can employ the Hellman-Feynman theorem.
For the linear Schr\"odinger equation, the slope of the level with respect 
to a change of a parameter, say $x_0$, is simply the expectation 
value (calculated with the help of the corresponding wavefunction)
of the derivative of the Hamiltonian with respect to $x_0$.
A diabatic passage
does not significantly change the wavefunction, so the change of
the slope is minimized. For the GPE, an analog of the Hellman-Feynman theorem
can be written as 
\begin{equation}
\frac{d\mu}{dx_0}=\langle\varphi|\frac{dV}{dx_0}+g\frac{d\varphi}{dx_0}\varphi^*+
g\varphi\frac{d\varphi^*}{dx_0}|\varphi\rangle,
\label{hf}
\end{equation}
and again if the change of the potential is so quick that the wavefunction does
not react significantly, the slope of the level remains the same.

\section{Summary and conclusions}

The proposed method of efficient collective excitation of a BEC seems
to be quite robust and allows preparing both solitonic and vortex-like
excitations of the condensate. Although the method is complementary to other
schemes, some of which have been already implemented experimentally,
still it may be advantageous in some cases. As we have discussed
already in \cite{kark01} our approach is somehow closest in spirit
to the adiabatic scheme proposed in \cite{dum98}. That approach seems
also to be quite robust and allows various excitations to be created.
The method of \cite{dum98} effectively uses two atomic internal states,
and 
thus involves a ``two-component'' condensate. Our approach
does not entangle internal and external excitations -- this may be
 advantageous in some applications. More importantly our diabatic
(or rather ``generalized diabatic'') method takes necessarily a short
time -- typically of the order of a few periods of the trap in our runs.
This has to be compared with several hundreds of periods necessary for
the excitation using the adiabatic process \cite{dum98}.

Needless to say while one may argue about the advantages of the proposed scheme
a most straightforward way to test it would be an experimental
approach. It seems that both the linear potential sweep for cigar shaped
BEC or the spiral like excitation for disc shaped quasi 2D condensates
require relatively minor changes in the already existing laboratory
set-ups.

We have found that ``dynamics'' of chemical potential in the time-independent 
GPE reveals interesting loop-like structures. This prevents us from
interpreting the changes of the chemical potential levels, 
with respect to the parameter, as some form of
level dynamics known from the linear Schr\"odinger equation. The observed 
``hysteresis-like''
behavior has its origin in the nonlinearity of the GPE  as explained
in the text. Its relation to the linear quantum mechanics of multiparticle
theory is being currently investigated. 

To summarize, we have proposed a simple scheme
 which enables us to create collective
excitations (both solitons and vortices)
of the Bose-Einstein condensate. This scheme may serve, we believe,
as an alternative to other proposed methods already utilized experimentally.

{\sl Note added in proof:} The scheme considered by us, similarly to other
works on this subject 
\cite{marzlin97,dum98,dobrek99,burger99,denschlag00,yukalov},
is based on the applicability of the GPE to the description of 
collectively excited
states of the BEC. That has been
 questioned very recently \cite{dziarmaga}.

\section{ACKNOWLEDGMENT}

We are grateful to  Maciek Lewenstein and Kazik Rz\c{a}\.zewski for
encouragement. Special thanks are due to Mariusz Gajda for suggesting the
imaginary time method to get stationary solutions of the GPE, and to Andrzej
Ostruszka for the help in solving computer problems.
Support of KBN under project 5~P03B~088~21 is acknowledged.
 
%%%%%references%%%%%%%%%%%%%%%%%%%%%%%%%%%%%%%%%%%%%%%%%%%%%%%%%%%%%%%%%%%%%%

\end{multicols}

\begin{references}
\bibitem{foot1} For simplicity we call by energy the value of
$\mu$ obtained for $\varphi(x)$ -- a time-independent solution of GPE.
Strictly speaking $\mu$ should be referred to as a chemical potential
while the energy per particle $\epsilon =\mu - g/2\int |\varphi(x)|^4dx$.
The plots of $\epsilon$ are similar to that of $\mu$ and are, 
therefore, not shown.
\end{references}
\end{document}